\documentclass[conference]{IEEEtran}
\IEEEoverridecommandlockouts
\usepackage{cite}
\usepackage{amsmath,amssymb,amsfonts}
\usepackage{algorithmic}
\usepackage{graphicx}
\usepackage{textcomp}
\usepackage{xcolor}

\usepackage[mode=tex]{standalone} 

\usepackage{amsmath,amssymb,amsfonts,bbm,bm,mathrsfs}
\usepackage{color}
\usepackage{algorithmic}
\usepackage{graphicx}
\usepackage{textcomp}
\usepackage{xcolor}
\usepackage{amsthm}

\usepackage[ruled,vlined,linesnumbered]{algorithm2e}

\usepackage{tikz}

\usepackage{chronosys}
\usepackage{xspace}
\usetikzlibrary{shapes,arrows,calc,positioning,fit,automata}
\usetikzlibrary{decorations.markings}
\usetikzlibrary{overlay-beamer-styles}
\usepackage{fontawesome}

\usepackage{pgfplots}
\usepackage{pgfplotstable}
\usepackage{filecontents}

\usepackage{caption,subcaption}

\newcommand{\Gcal}{\mathcal{G}}
\newcommand{\Fcal}{\mathcal{F}}

\newcommand{\E}{\mathbb{E}}
\newcommand{\funF}{\mathrm{f}}
\newcommand{\funC}{\mathrm{c}}
\newcommand{\funQ}{\mathrm{q}}
\newcommand{\funG}{\mathrm{g}}

\newcommand{\mat}[1]{\ensuremath{\mathbf{#1}}}
\newcommand{\vect}[1]{\ensuremath{\mathbf{#1}}}

\newcommand{\Norm}[1]{\left\lVert#1\right\rVert}

\newcommand{\Pilots}{\mat{\tilde{X}}}
\newcommand{\pilots}{\vect{\tilde{x}}}

\newcommand{\NoisePilots}{\mat{Z}}
\newcommand{\noisePilots}{\vect{z}}

\newcommand{\myBlue}{blue!80!black}
\newcommand{\myGreen}{green!60!black}
\newcommand{\myRed}{red!80!black}

\definecolor{NYUviolet}{HTML}{57068c} 	
\definecolor{NYUlight}{HTML}{8900e1} 	
\definecolor{NYUdark}{HTML}{330662} 	
\definecolor{NYUnight}{HTML}{220337} 	

\tikzstyle{block}=[rectangle,draw,very thick,fill=white,align=center,font=\Large]
\tikzstyle{edge} = [draw,very thick,->,-triangle 45]
\tikzstyle{plateBlue} = [draw=\myBlue, shape=rectangle, rounded corners=0.5ex, ultra thick,
minimum width=3.1cm, text=\myBlue, text width=3.1cm, align=right, inner sep=10pt, inner ysep=10pt,append after command={node[font=\bf,text=\myBlue,above left= 3pt of \tikzlastnode.south east] {#1}}]

\tikzstyle{plateRed} = [draw=\myRed, shape=rectangle, rounded corners=0.5ex, ultra thick,
minimum width=3.1cm, text=\myRed, text width=3.1cm, align=right, inner sep=10pt, inner ysep=10pt,append after command={node[font=\bf\LARGE,text=\myRed,above= 3pt of \tikzlastnode.south] {#1}}]

\tikzstyle{plateGreen} = [draw=\myGreen, shape=rectangle, rounded corners=0.5ex, ultra thick,
minimum width=3.1cm, text=green!80!black, text width=3.1cm, align=right, inner sep=10pt, inner ysep=10pt,append after command={node[font=\bf\LARGE,text=\myGreen,above= 3pt of \tikzlastnode.south] {#1}}]

\tikzstyle{plate} = [draw, shape=rectangle, rounded corners=0.5ex, ultra thick,
minimum width=3.1cm,  text width=3.1cm, align=right, inner sep=10pt, inner ysep=10pt,append after command={node[font=\bf\Large,above left= 3pt of \tikzlastnode.south east] {#1}}]

\tikzstyle{plateSmall} = [draw, shape=rectangle, rounded corners=0.5ex, ultra thick,
minimum width=3.1cm,  text width=3.1cm, align=right, inner sep=10pt, inner ysep=10pt,append after command={node[font=\bf,above= 3pt of \tikzlastnode.south] {#1}}]

\usepackage[prefix=]{xcolor-material}
\pgfmathdeclarerandomlist{material}{%
	{Red}{Blue}{Green}}
\tikzset{%
	half clip/.code={
		\clip (0, -256) rectangle (256, 256);
	},
	color/.code=\colorlet{fill color}{#1},
	color alias/.code args={#1 as #2}{\colorlet{#1}{#2}},
	colors alias/.style={color alias/.list/.expanded={#1}},
	execute/.code={#1},
	on left/.style={.. on left/.style={#1}},
	on right/.style={.. on right/.style={#1}},
	split/.style args={#1 and #2}{
		on left ={color alias=fill color as #1},
		on right={color alias=fill color as #2, half clip}
	}
}
\newcommand\reflect[2][]{%
	\begin{scope}[#1]\foreach \side in {-1, 1}{\begin{scope}
				\ifnum\side=-1 \tikzset{.. on left/.try}\else\tikzset{.. on right/.try}\fi
				\begin{scope}[xscale=\side]#2\end{scope}
\end{scope}}\end{scope}}

\tikzset{%
	cat/.pic={
		\tikzset{x=1.5cm/5,y=1.5cm/5,shift={(0,-1/3)}}
		\useasboundingbox (-1,-1) (1,2);
		\fill [BlueGrey900] (0,-2)
		.. controls ++(180:3) and ++(0:5/4) .. (-2,0)
		arc (270:90:1/5)
		.. controls ++(0:2) and ++(180:11/4) .. (0,-2+2/5);
		\foreach \i in {-1,1}
		\scoped[shift={(1/2*\i,9/4)}, rotate=45*\i]{
			\clip [overlay] (0, 5/9) ellipse [radius=8/9];
			\clip [overlay] (0,-5/9) ellipse [radius=8/9];
			\fill [BlueGrey900] ellipse [radius=1];
			\clip [overlay] (0, 7/9) ellipse [radius=10/11];
			\clip [overlay] (0,-7/9) ellipse [radius=10/11];
			\fill [Purple100] ellipse [radius=1];
		};
		\fill [BlueGrey900] ellipse [x radius=3/4, y radius=2];
		\fill [BlueGrey100] ellipse [x radius=1/3, y radius=1];
		\fill [BlueGrey900]
		(0,15/8) ellipse [x radius=1, y radius=5/6]
		(0, 8/6) ellipse [x radius=1/2, y radius=1/2]
		{[shift={(-1/2,-2)}, rotate= 10]  ellipse [x radius=1/3, y radius=5/4]}
		{[shift={( 1/2,-2)}, rotate=-10] ellipse [x radius=1/3, y radius=5/4]};
		\fill [BlueGrey500]
		(-1/9,11/8) ellipse [x radius=1/5, y radius=1/5]
		( 1/9,11/8) ellipse [x radius=1/5, y radius=1/5];
		\fill [Purple100]
		(0,12/8)     ellipse [x radius=1/10, y radius=1/5]
		(0,12/8+1/9) ellipse [x radius=1/5 , y radius=1/10];
		\foreach \i in {-1,1}
		\scoped[shift={(1/2*\i,2)}, rotate=35*\i]{
			\clip [overlay] (0, 1/7) ellipse [radius=2/7];
			\clip [overlay] (0,-1/7) ellipse [radius=2/7];
			\fill [Yellow50] ellipse [radius=1];
		};
		\scoped{
			\clip (-1,-2) rectangle ++(2,1);
			\fill [BlueGrey900] (0,-2) ellipse [radius=1/2];
			\fill [Grey100]
			(-1/2,-2) ellipse [x radius=1/3, y radius=1/4]
			( 1/2,-2) ellipse [x radius=1/3, y radius=1/4];
		};
		\foreach \i in {-1,1}
		\foreach \j in {-1,0,1}
		\fill [Grey100, shift={(0,11/8)}, xscale=\i, rotate=\j*15,
		shift=(0:1/2)]
		ellipse [x radius=1/3, y radius=1/64];
	},
dog/.pic={
	\begin{scope}[x=1.5cm/480,y=1.5cm/480]
		\useasboundingbox (-256, -256) (256, 256);
		\reflect[split=Brown400 and Brown500]{
			\fill [fill color] (0,-64) ellipse [x radius=160, y radius=144];
			\fill [fill color] (0, 32) ellipse [x radius=128, y radius=112];
			\fill [fill color] (32,96)
			.. controls ++( 75:128) and ++(105:128) .. ++(192,  0)
			.. controls ++(285: 96) and ++(285: 96) .. ++(-80,-32)
			.. controls ++(105: 64) and ++( 75: 32) .. cycle;
		}
		\reflect[split={Grey100 and Grey200}]{
			\clip (0,-64) ellipse [x radius=160, y radius=144];
			\fill [fill color](0,-224) 
			.. controls ++(  0:64) and ++(270:64) .. ++(96,64)
			.. controls ++( 90:64) and ++(270:64) .. ++(-96,192)
			.. controls ++(270:64) and ++( 90:64) .. ++(-96,-192)
			.. controls ++(270:64) and ++(180:64) .. cycle;
		}
		\reflect[split={Pink100 and Pink200}]{
			\fill [fill color](0,-192) ellipse [x radius=28, y radius=32];
		}
		\reflect[split={BlueGrey800 and BlueGrey900}]{
			\fill [fill color](0,-144) 
			.. controls ++(  0:20) and ++(315:20) .. ++( 40,64)
			.. controls ++(135:20) and ++( 45:20) .. ++(-80, 0)
			.. controls ++(225:20) and ++(180:20) .. cycle;
			\fill [BlueGrey900] (56, 0) circle [radius=20];
			\fill [fill color] (-8,-112)
			-- ++(16,0) -- ++(0,-32) arc (180:360:24)
			arc (180:0:8) arc (360:180:40);
		}
\end{scope}}
}

\tikzset{
	o/.style={
		shorten >=#1,
		decoration={
			markings,
			mark={
				at position 1
				with {
					\draw circle [radius=#1];
				}
			}
		},
		postaction=decorate
	},
	o/.default=2pt
}

\tikzset{naming/.style={align=center,font=\large}}
\tikzset{antenna/.style={insert path={-- coordinate (ant#1) ++(0,0.25) -- +(135:0.25) + (0,0) -- +(45:0.25)}}}
\tikzset{station/.style={naming,draw,shape=dart,shape border rotate=90, minimum width=15mm, minimum height=30mm,outer sep=0pt,inner sep=3pt}}
\tikzset{stationPoster/.style={naming,draw,shape=dart,shape border rotate=90, minimum width=20mm, minimum height=40mm,outer sep=0pt,inner sep=3pt}}
\tikzset{mobile/.style={naming,draw,shape=rectangle,minimum width=12mm,minimum height=6mm, outer sep=0pt,inner sep=3pt}}
\tikzset{radiation/.style={{decorate,decoration={expanding waves,angle=90,segment length=4pt}}}}

\def\BibTeX{{\rm B\kern-.05em{\sc i\kern-.025em b}\kern-.08em
    T\kern-.1667em\lower.7ex\hbox{E}\kern-.125emX}}
\begin{document}

\title{
    \huge{
        Precoding-oriented Massive MIMO CSI Feedback Design
    }
    \thanks{
        This work was done in part while F. Carpi was an intern at Nokia Bell Labs.
        The work of F.~Carpi, S.~Garg, and E.~Erkip was supported in part by the NYU WIRELESS Industrial Affiliates Program, by the NSF--Intel grant \#2003182, and by the NSF grant \#1925079.
        Please send correspondence to \texttt{fabrizio.carpi@nyu.edu}.
    }
}

\author{
    \IEEEauthorblockN{
        Fabrizio Carpi\IEEEauthorrefmark{1}, 
        Sivarama Venkatesan\IEEEauthorrefmark{2}, 
        Jinfeng Du\IEEEauthorrefmark{2}, 
        Harish Viswanathan\IEEEauthorrefmark{2}, 
        Siddharth Garg\IEEEauthorrefmark{1}, 
        Elza Erkip\IEEEauthorrefmark{1}}
    \IEEEauthorblockA{
        \IEEEauthorrefmark{1}Department of Electrical and Computer Engineering, New York University, Brooklyn, NY\\
        \IEEEauthorrefmark{2}Nokia Bell Labs, Murray Hill, NJ}
}

\maketitle

\begin{abstract}
Downlink massive multiple-input multiple-output (MIMO) precoding algorithms in frequency division duplexing (FDD) systems rely on accurate channel state information (CSI) feedback from users.
In this paper, we analyze the tradeoff between the CSI feedback overhead and the performance achieved by the users in systems in terms of achievable rate.
The final goal of the proposed system is to determine the beamforming information (i.e., precoding) from channel realizations.
We employ a deep learning-based approach to design the end-to-end precoding-oriented feedback architecture, that includes learned pilots, users' compressors, and base station processing.
We propose a loss function that maximizes the sum of achievable rates with minimal feedback overhead.
Simulation results show that our approach outperforms previous precoding-oriented methods, and provides more efficient solutions with respect to conventional methods that separate the CSI compression blocks from the precoding processing.
\end{abstract}



\section{Introduction}
\label{sec:introduction}

Massive multiple-input multiple-output (MIMO) technology became one of the pillars of 5G wireless systems and beyond, and accurate channel state information (CSI) is a key enabler to unlock its full potential~\cite{mMIMO-tenMyths}.
Driving into 6G wireless systems, with even larger antenna arrays and bandwidths, effective CSI becomes a crucial requirement to achieve the desired performance.  
When adopting time division duplexing (TDD), the base station (BS) exploits channel reciprocity to obtain CSI through uplink transmissions.
On the other hand, channel reciprocity does not hold when considering frequency division duplexing (FDD).
In FDD, the BS needs an accurate CSI in order to serve users with high spectral efficiency.
In FDD systems, the users estimate the downlink channel realizations and feed back the estimated CSI on the uplink.
The transmission of the downlink CSI on the uplink constitutes a communication overhead.
As the system dimension increases (e.g., more antennas, more users), the communication cost of the feedback overhead becomes prohibitive. 

Classical CSI compression techniques make use of vector quantization~\cite{Love-2008} and compressed sensing~\cite{Rao-2014}:  in the first case, the overhead is still impractical for large systems, while the latter technique assumes channel sparsity in some domains.
Overall, three fundamental metrics can be considered for the CSI feedback problem: \emph{overhead}, that is the number of bits sent on the feedback link; \emph{performance}, that is the total achievable rate at the users; \emph{distortion}, that is the loss (e.g., mean squared error) incurred when trying to reconstruct the channel realizations at the BS.
Generally speaking, it is known that the optimal rate is achieved with nonlinear precoding methods~\cite{Caire2003, PrecodingMUMIMO} based on dirty paper coding~\cite{Costa1983}.
In this work, we focus on the sum of achievable rates under the assumption that transmit beamforming is implemented with linear precoding~\cite{MU-beamforming}.

Recently, deep learning-based solutions have been proposed for the CSI feedback problem in massive MIMO FDD systems, see~\cite{DL-CSI-FB-survey-2022} and references therein for an overview.
In general, these data-driven solutions rely on fewer assumptions and outperform classical CSI feedback methods.
The capability of the autoencoder~\cite{Goodfellow-et-al-2016} structure for the CSI compression problem was first shown by~\cite{DL-CSI-FB-2018}.
Several follow-up works have improved the distortion metric~\cite{DL-CSI-FB-survey-2022} with increasing dimensionality reduction on the feedback link.
Similarly to image processing applications~\cite{Balle-2017, Balle-2018}, a mechanism for the feedback overhead optimization has been introduced in~\cite{ICL-2021}. 
The authors~\cite{ICL-2021} consider the rate-distortion objective, where the goal is to reconstruct channel realizations with minimal overhead at the BS side.
Their results~\cite{ICL-2021} show that the feedback overhead can be further reduced with respect to previous work, but there is no focus on the final task that is performed at the BS (e.g., beamforming).
On the other hand, \cite{Toronto-2021, implicit-CSI-2022}~consider objectives related to the final task to be performed at the BS, i.e., beamforming with linear precoding.
A single-user system is analyzed in~\cite{implicit-CSI-2022}, where precoding information is computed and compressed at the user side, and decoded at the base station: the proposed solution shows overhead savings compared with 3GPP standards methods.
Instead, \cite{Toronto-2021}~proposes a multi-user precoder-oriented architecture that learns pilots, user processing, and BS processing from training data.
The system output is the collection of linear precoders and the objective function is the sum of achievable rates experienced by the users.
Their best results~\cite[Fig. 9]{Toronto-2021} are obtained by modeling each feedback tap as a binary value, using a smooth approximation during training to allow backpropagation.
Hence, the feedback overhead is determined by the choice of the feedback dimension (fixed), without the possibility to obtain further compression.

In this paper, we analyze the tradeoff between the CSI feedback overhead and the system performance in terms of the sum of achievable rates experienced by the users. 
In general, the CSI is compressed and sent over a rate-limited link by the users, and the BS has to process the CSI feedback in order to determine the beamforming information.
We assume that linear precoding is adopted, and we focus on a task-oriented approach: the final task for the BS is to determine precoding vectors for each user, and the system's goal is to maximize the sum of achievable rates.
We adopt a multi-user precoding-oriented end-to-end architecture similar to~\cite{Toronto-2021}, where the users observe a sequence of noisy pilots as an input, and produce precoding-oriented feedback messages for the BS. 
The BS processes the received feedback and determines the precoder vectors for each user.
The pilots, feedback schemes, and BS processing are modeled with neural networks.
Differently from~\cite{Toronto-2021}, we include a feedback overhead optimization mechanism that estimates the feedback distribution during training and uses entropy coding to generate the bits streams at test time~\cite{Balle-2017,Balle-2018}. 
The entropy of the feedback values is used to estimate the feedback overhead (feedback rate).
The same feedback optimization method~\cite{Balle-2017, Balle-2018} was used in~\cite{ICL-2021}, but in the context of CSI feedback for channel reconstruction without including pilots design.
In order to train the end-to-end architecture with gradient descent, we propose a tunable loss function that captures the tradeoff between feedback overhead and system performance.
We show that the precoding-oriented system trained over the overhead-performance loss function outperforms conventional methods based on channel reconstruction followed by traditional precoding techniques.
In our precoding-oriented approach, the user network is able to learn how to efficiently transfer precoding-oriented quantized information over the feedback link.

We illustrate the system model in Section~\ref{sec:system-model}, while the precoding-oriented CSI feedback approach is discussed in Section~\ref{sec:SemCom-CSI}.
Numerical results are shown in Section~\ref{sec:results}, while conclusions are drawn in Section~\ref{sec:conclusion}.

\section{System Model}
\label{sec:system-model}

We consider a massive MIMO system where a BS with $N_t$ transmit antennas serves $K$ single-antenna users.
We assume that linear precoding is used at the BS, hence the downlink transmitted signal is
\begin{equation}
    \vect{x} = \sum_{k=1}^K \vect{v}_k s_k = \vect{V}\vect{s}
\end{equation}
where $\vect{v}_k\in\mathbb{C}^{N_t}$ is the precoding vector and $s_k\in\mathbb{C}$ is the symbol to be sent by the $k$-th user.
The precoding matrix $\vect{V}\in\mathbb{C}^{N_t\times K}$ satisfies the total power constraint $\text{Tr}(\vect{V}\vect{V}^H) \leq P$, and the symbols $\vect{s}\in\mathbb{C}^K$ are normalized to $\mathbb{E}[\vect{s}\vect{s}^H] = \vect{I}$.
The received signal at the $k$-th user is 
\begin{equation}
\label{eq:rx-signal}
    y_k = \vect{h}_k^H \vect{v}_k s_k + \sum_{j\neq k} 
          \vect{h}_k^H \vect{v}_j s_j + 
          z_k
\end{equation}
where $\vect{h}_k\in\mathbb{C}^{N_t}$ is the vector of downlink channel gains for the $k$-th user and $z_k\sim \mathcal{CN}(0,\sigma^2)$ is the additive white Gaussian noise.
The matrix containing the channel gains for all the users is denoted by $\mat{H}\in\mathbb{C}^{N_t\times K}$.
Given the received signal model in~\eqref{eq:rx-signal}, the achievable rate for the $k$-th user is 
\begin{equation}
\label{eq:ach-rate-user}
    R_k = \log_2 \left( 1 + \frac{ |\vect{h}_k^H \vect{v}_k|^2 }{ \sum_{j\neq k } |\vect{h}_k^H \vect{v}_j|^2 + \sigma^2} \right).
\end{equation}
Considering the multi-terminal communication problem with $K$ users, the performance can be expressed as the sum of achievable rates, i.e., 
\begin{equation}
\label{eq:ach-rate-sum}
    R = \sum_{k=1}^K R_k.
\end{equation}

Accurate CSI is crucial to design a precoding matrix $\mat{V}$ that maximizes the sum rate in~\eqref{eq:ach-rate-sum}.
We assume that both the BS and the users do not have a-priori knowledge about the channel realizations.
Hence, the BS has to learn the downlink precoding matrix $\mat{V}$ based on the feedback received from the users on the uplink.
Therefore, we assume that the BS sends reference signals (pilots) during the downlink training phase.
The pilots, of length $L$, are denoted by $\Pilots\in\mathbb{C}^{N_t\times L}$.
The received noisy pilots $\vect{\tilde{y}}_k\in\mathbb{C}^{L}$ at $k$-th user are
\begin{equation}
\label{eq:pilots-noisy}
    \vect{\tilde{y}}_k = \vect{h}_k^H \Pilots + \vect{z}_k,
\end{equation}
where $\vect{z}_k\sim\mathcal{CN}(\vect{0},\sigma^2\vect{I})$ is the additive white Gaussian noise at the $k$-th user.
The $\ell$-th pilot transmission satisfies the instantaneous power constraint $\Norm{\pilots_\ell}_2^2 \leq P$, where $\vect{\tilde{x}}_\ell$ is the $\ell$-th column of $\vect{\tilde{X}}$.

In general, the CSI feedback overhead is proportional to the dimensions of the system (e.g., number of users, antennas), and becomes very large in case of massive MIMO systems with many users.
Hence, the users are required to extract relevant CSI from the received pilots, then feed back the compressed CSI, or other relevant information needed for the downlink precoding, to the BS over a rate-limited link.
The feedback $\vect{b}_k$ for the $k$-th user is denoted by
\begin{equation}
\label{eq:feedback-k}
    \vect{b}_k = \Fcal( \vect{\tilde{y}}_k ),
\end{equation}
where $\mathcal{F}$ represents the feedback scheme adopted by the users.
We focus on the regime where the feedback overhead required to transmit $\vect{b}_k$ is much smaller than sending the channel realizations explicitly.

The BS collects the received feedback from all the $K$ users, denoted by $(\vect{b_1},\dots,\vect{b}_K)$, and outputs both the precoding matrix $\mat{V}$ and the reconstructed channel matrix $\mat{\hat{H}}$:
\begin{equation}
\label{eq:BS-output}
    (\mat{V}, \mat{\hat{H}}) = \Gcal (\vect{b_1},\dots,\vect{b}_K),
\end{equation}
where $\mathcal{G}$ denotes the BS processing.
Note that in~\eqref{eq:BS-output} both the precoders $\mat{V}$ and the channel reconstructions $\mat{\hat{H}}$ are potentially provided by the BS. 
However, in this paper we focus on the precoding-oriented approach, i.e., only $\mat{V}$ is provided at the output of the BS, as the ultimate task of maximizing sum rate in~\eqref{eq:ach-rate-sum} only depends on $\mat{V}$.
We use the reconstructions $\mat{\hat{H}}$ to compare our method with traditional techniques based on the separation between CSI feedback for channel reconstruction~\cite{DL-CSI-FB-2018,ICL-2021,DL-CSI-FB-survey-2022} and conventional precoding techniques.

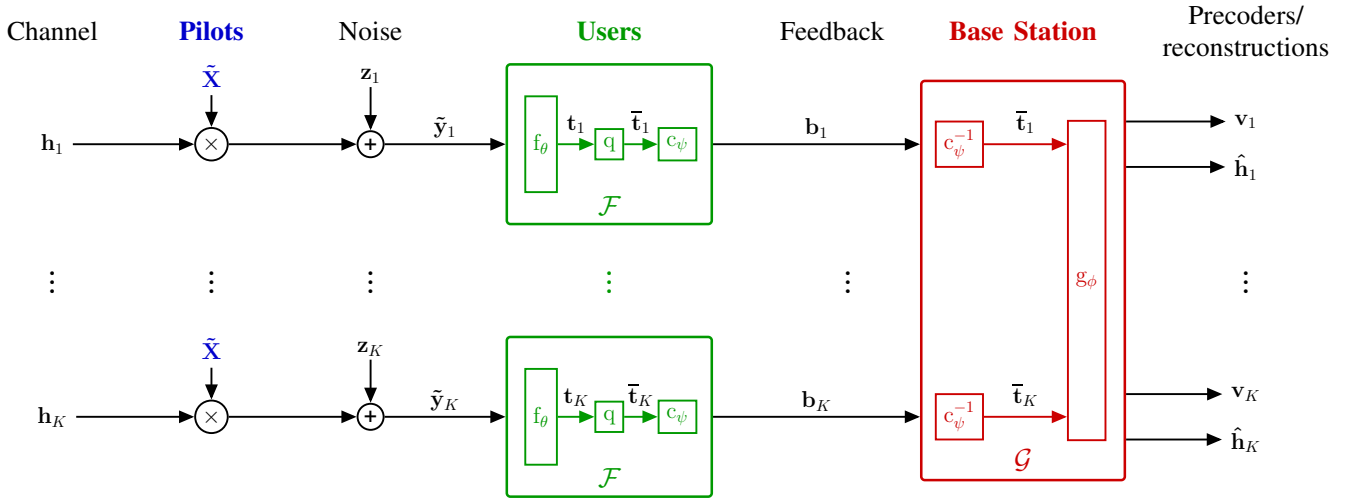
\begin{figure*}[t]
	\centering
	\begin{tikzpicture}
	\def\dhor{7}
	\def\dUE{1.5}
	\def\dvert{6}
	\def\dout{0.5}
	\def\dz{1.5}
	\def\dhead{2.5}

	\node[block, draw=white, inner sep=5pt] (H1) at (-1.75*\dhor,0) {$\vect{h}_1$};
	\node[block, draw=white, inner sep=5pt] (HK) at (-1.75*\dhor,-\dvert) {$\vect{h}_K$};
	\node[block, draw=white, inner sep=5pt, rotate=90] (dots) at (-1.75*\dhor,-0.5*\dvert) {$\textbf{\dots}$};
	\node[font=\LARGE] () at (-1.75*\dhor,+\dhead) {Channel};
	
	\node[block, inner sep=2pt, align=center, circle] (Pilots1) at (-1.25*\dhor,0) {$\times$};
	\node[block, inner sep=2pt, align=center, circle] (PilotsK) at (-1.25*\dhor,-\dvert) {$\times$};
	\node[block, draw=white, rotate=90] (dots) at (-1.25*\dhor,-0.5*\dvert) {$\textbf{\dots}$};
    \node[text=\myBlue, font=\Large, inner sep=5pt, align=center] (X1) at (-1.25*\dhor,\dz) {$\Pilots$};
	\node[text=\myBlue, font=\Large, inner sep=5pt, align=center] (XK) at (-1.25*\dhor,-\dvert+\dz) {$\Pilots$};
	\node[font=\LARGE\bf, text=\myBlue] () at (-1.25*\dhor,+\dhead) {Pilots};
	
	\node[block, inner sep=2pt, align=center, circle] (plus1) at (-0.75*\dhor,0) {\textbf{+}};
	\node[block, inner sep=2pt, align=center, circle] (plusK) at (-0.75*\dhor,-\dvert) {\textbf{+}};
	\node[block, draw=white, rotate=90] (dots) at (-0.75*\dhor,-0.5*\dvert) {$\textbf{\dots}$};

	\node[block, font=\Large, inner sep=2pt, align=center, draw=white] (noise1) at (-0.75*\dhor,\dz) {$\vect{z}_1$};
	\node[block, font=\Large, inner sep=2pt, align=center, draw=white] (noiseK) at (-0.75*\dhor,\dz-\dvert) {$\vect{z}_K$};
	\node[font=\LARGE] () at (-0.75*\dhor,+\dhead) {Noise};

	\node[block, text=\myGreen, draw=\myGreen, inner sep=5pt, align=center, minimum height=60] (enc1) at (-\dUE,0) {$\funF_\theta$};
	\node[block, text=\myGreen, draw=\myGreen, inner sep=5pt, align=center] (quant1) at (0,0) {$\funQ$};
	\node[block, text=\myGreen, draw=\myGreen, inner sep=5pt, align=center] (entcod1) at (\dUE,0) {$\funC_\psi$};
	\node[plateGreen=$\mathcal{F}$, fit=(enc1)(quant1)(entcod1),  minimum width=50, minimum height=100] (UE1) at (0,0) {};
	\node[font=\LARGE\bf, text=\myGreen] () at (0,+\dhead) {Users};

	\node[block, text=\myGreen, draw=\myGreen, inner sep=5pt, align=center, minimum height=60] (encK) at (-\dUE,-\dvert) {$\funF_\theta$};
	\node[block, text=\myGreen, draw=\myGreen, inner sep=5pt, align=center] (quantK) at (0,-\dvert) {$\funQ$};
	\node[block, text=\myGreen, draw=\myGreen, inner sep=5pt, align=center] (entcodK) at (\dUE,-\dvert) {$\funC_\psi$};
	\node[plateGreen=$\mathcal{F}$, fit=(encK)(quantK)(entcodK),  minimum width=50, minimum height=100] (UEK) at (0,-\dvert)  {};
	
	\node[block, text=\myGreen, draw=white, rotate=90, inner sep=0] (dots) at (0,-0.5*\dvert) {$\textbf{\dots}$};
	
	\node[block, draw=white, inner sep=10pt, rotate=90] (dots) at (0.75*\dhor,-0.5*\dvert) {$\textbf{\dots}$};
	
	\node[block, text=\myRed, draw=\myRed, inner sep=5pt] (entdecK) at (1.1*\dhor,-1*\dvert) {$\funC_\psi^{-1}$};
	\node[block, text=\myRed, draw=\myRed, inner sep=5pt] (entdec1) at (1.1*\dhor,0) {$\funC_\psi^{-1}$};
	\node[block, text=\myRed, draw=\myRed, inner sep=5pt, minimum height=200] (dec) at (1.5*\dhor,-0.5*\dvert) {$\funG_\phi$};
	\node[plateRed=$\mathcal{G}$, fit=(entdecK)(dec),  minimum width=50, minimum height=250] (BS) at (1.3*\dhor,-0.5*\dvert) {};
	\node[text=\myRed, font=\bf\LARGE] () at (1.3*\dhor,+\dhead) {Base Station};
	
	\node[block, draw=white, inner sep=5pt] (Hhat1) at (2*\dhor,-\dout) {$\vect{\hat{h}}_1$};
	\node[block, draw=white, inner sep=5pt] (HhatK) at (2*\dhor,-\dout-\dvert) {$\vect{\hat{h}}_K$};
	\node[block, draw=white, inner sep=5pt] (V1) at (2*\dhor,+\dout) {$\vect{v}_1$};
	\node[block, draw=white, inner sep=5pt] (VK) at (2*\dhor,+\dout-\dvert) {$\vect{v}_K$};
	\node[block, draw=white, inner sep=5pt, rotate=90] (dots) at (2*\dhor,-0.5*\dvert) {$\textbf{\dots}$};
	
	\node[font=\LARGE] () at (0.7*\dhor,+\dhead) {Feedback};
	
	\node[font=\LARGE, align=center] () at (2*\dhor,+\dhead) {Precoders/\\reconstructions};


	\path (H1) edge[edge] (Pilots1);
	\path (HK) edge[edge] (PilotsK);
	
	\path (X1) edge[edge] (Pilots1);
	\path (XK) edge[edge] (PilotsK);
	
	\path (Pilots1) edge[edge] (plus1);
	\path (PilotsK) edge[edge] (plusK);
	
	
	\path (noise1) edge[edge] (plus1);
	\path (noiseK) edge[edge] (plusK);
	
	\path (plus1) edge[edge] node[above, font=\Large]{$\vect{\tilde{y}}_1$} (UE1);
	\path (plusK) edge[edge] node[above, font=\Large]{$\vect{\tilde{y}}_K$} (UEK);

	\path[edge] (UE1.east) --node[above,font=\Large]{$\vect{b}_1$} (UE1-|BS.west);
	\path (UEK.east) edge[edge] node[above,font=\Large] {$\vect{b}_K$} (UEK-|BS.west);
	
	\path (BS.east|-V1.west) edge[edge] (V1.west);
	\path (BS.east|-VK.west) edge[edge] (VK.west);
	
	\path (BS.east|-Hhat1.west) edge[edge] (Hhat1.west);
	\path (BS.east|-HhatK.west) edge[edge] (HhatK.west);
	
	
	\path[edge, draw=\myGreen] (enc1) --node[above=0.1,font=\Large]{$\vect{t}_1$} (quant1);
	\path[edge, draw=\myGreen] (quant1) --node[above=0.1,font=\Large]{$\vect{\overline{t}}_1$} (entcod1);
	\path[edge, draw=\myRed] (entdec1.east) --node[above=0.1,font=\Large]{$\vect{\overline{t}}_1$} (entdec1-|dec.west);
	
	\path[edge, draw=\myGreen] (encK) --node[above=0.1,font=\Large]{$\vect{t}_K$} (quantK);
	\path[edge, draw=\myGreen] (quantK) --node[above=0.1,font=\Large]{$\vect{\overline{t}}_K$} (entcodK);
	\path[edge, draw=\myRed] (entdecK.east) --node[above=0.1,font=\Large]{$\vect{\overline{t}}_K$} (entdecK-|dec.west);

\end{tikzpicture}
	\caption{System model.
	The pilots (marked in blue) are learned during the end-to-end learning.
	The green boxes correspond to the users' feedback scheme $\mathcal{F}$, which is composed of the DNN $\funF_\theta$, the quantizer $\funQ$ and the entropy coder $\funC_\psi$.
	The BS processing is denoted with $\mathcal{G}$, which is composed by the entropy decoder $\funC^{-1}_\psi$ and the DNN $\funG_\phi$.
	The output of the BS processing are the precoders $\vect{v}_k$ or the channel reconstructions $\vect{\hat{h}}_k$ depending on whether the precoding-oriented approach is adopted or not.
	}
	\label{fig:system-model}
\end{figure*}

Our focus in this paper is a precoding-oriented architecture, where the tradeoff between the feedback overhead and the users' performance is considered.
Our system seeks to maximize the sum of achievable rates~\eqref{eq:ach-rate-sum}, while the amount of bits required to transmit $(\vect{b}_1,\dots,\vect{b}_K)$ over the feedback link is bounded.
We use neural networks to design the pilots $\Pilots$, the feedback scheme $\mathcal{F}$, and the BS processing $\mathcal{G}$.
The overall system model is shown in Fig.~\ref{fig:system-model}. 
More details about each processing block are provided in Section~\ref{sec:SemCom-CSI}.

\section{Precoding-oriented CSI Feedback with Overhead-performance Tradeoff}
\label{sec:SemCom-CSI}

Consider the block diagram in Fig.~\ref{fig:system-model}.
Similarly to~\cite{Toronto-2021}, we use neural networks in place of conventional methods for the pilots $\Pilots$, the feedback scheme $\Fcal$, and the BS scheme $\Gcal$.
We also adopt a mechanism, proposed in~\cite{Balle-2017, Balle-2018}, that optimizes the compression and quantization of the feedback.
A similar approach was also used in~\cite{ICL-2021} in the context of CSI feedback; however, the goal in that case was to reconstruct the channel coefficients, while we focus on the design of the precoder vectors.
We propose a precoding-oriented loss function, where the overhead-performance tradeoff is directly embedded in the objective function.
In this way, the feedback scheme $\Fcal$ can learn an efficient precoding-oriented representation of the channel realization, and the BS processing $\Gcal$ is able to directly output the precoding matrix $\mat{V}$.
More details about each processing block are provided in the next sections.

\subsection{Downlink Pilots}
\label{sec:pilots}

The downlink received (noisy) pilots for the $k$-th user are expressed in~\eqref{eq:pilots-noisy}.
As in~\cite{Toronto-2021}, we model the pilots $\Pilots$ as a fully-connected neural network layer with linear activation and zero bias.
The power constraint $P$ is guaranteed during training by setting $\Norm{\pilots_\ell}_2^2 = P$. 
Gaussian noise $\vect{z}_k$ is added to the sequence of $L$ received pilots to model the receiver noise according to~\eqref{eq:pilots-noisy}.

\subsection{Feedback Scheme}
\label{sec:feedback}

As depicted in Fig.~\ref{fig:system-model}, the feedback scheme for each user is denoted by $\mathcal{F}$.
We assume that $\mathcal{F}$ is the same scheme for all the $K$ users, while each terminal observes different channel realizations $\vect{h}_1,\dots,\vect{h}_K$.
As observed in~\cite{Toronto-2021,ICL-2021}, using a unique $\mathcal{F}$ works well also in the multi-user scenario if the channel realizations have the same statistics according to the channel model.
The feedback scheme is composed of three components: a deep neural network (DNN) $\funF_\theta$, a quantizer $\funQ$, and an entropy coder $\funC_\psi$.
More details about the latter two are also provided in the next sections.

The DNN $\funF_\theta$, where $\theta$ represents the set of trainable parameters, is used to extract features from the received pilots $\vect{\tilde{y}}_k$. 
The output of the DNN is denoted as $\vect{t}_k = \funF_\theta(\vect{\tilde{y}}_k)$, $\vect{t}_k\in\mathbb{R}^{N_b}$.
Similarly to~\cite{Toronto-2021}, we consider a fully-connected DNN.
More details about the DNN architecture are provided in Section~\ref{sec:results}.

The quantizer $\funQ$ performs uniform scalar quantization to the closest integer.
The quantized vector for the $k$-th user is denoted as $\vect{\overline{t}}_k = \funQ(\vect{t}_k)$.
During training, the quantization is replaced by adding independent identically distributed (iid) uniform noise $\vect{u}_k$, where the width of the uniform distribution is equal to the quantization bin width, i.e.,  $u_k^{1},\dots,u_k^{N_b}\sim\mathcal{U}[-0.5,+0.5]$~\cite{Balle-2017}. We denote the \emph{pseudo-quantized} vector as $\vect{\tilde{t}}_k = \vect{t}_k + \vect{u}_k$ and it  substitutes $\vect{\overline{t}}_k$ during training to allow gradient backpropagation.

The entropy coder $\funC_\psi$, where $\psi$ represents the set of trainable parameters, converts the quantized vector $\vect{\overline{t}}_k$ into bit streams in a lossless fashion. 
As in~\cite{Balle-2018}, $\vect{\tilde{t}}_k$ is modeled using a parametric, fully factorized density function.
Each element of $\vect{\tilde{t}}_k$ is modeled as a zero-mean Gaussian distribution with standard deviation learned during training. 
These learned parameters are then used by $\funC_\psi$ to encode $\vect{\overline{t}}_k$ at test time.

\subsection{Feedback Overhead Optimization}
\label{sec:feedback-overhead}

Similarly to previous works in image processing~\cite{Balle-2017,Balle-2018} and CSI feedback~\cite{ICL-2021}, we consider the feedback rate as part of our optimization objective.
Since both the quantizer and the entropy coder are not differentiable functions, they are substituted by iid uniform noise during training~\cite{Balle-2017}, as described above.
The iid noise $\vect{u}_k$ simulates the quantization noise.
The compression performed by the entropy coder is lossless and at a rate close to the entropy of $\vect{\overline{t}}_k$; so, at the BS side we have $\funC_\psi^{-1}(\vect{b}_k) = \vect{\overline{t}}_k$.
In fact, during training, the entropy of $\vect{\tilde{t}}_k$ is estimated in terms of the model parameters $\psi$.
Note that the probability density of $\vect{\tilde{t}}_k$ is a continuous relaxation of the probability mass function of $\vect{\overline{t}}_k$~\cite{Balle-2017}, hence the differential entropy of $\vect{\tilde{t}}_k$ approximates the entropy of  $\vect{\overline{t}}_k$; the estimated entropy represents the average bit rate at the quantizer output and will be used in our loss function to measure the feedback overhead~\cite{Balle-2017}.
During testing, the noise $\vect{u}_k$ is not injected, but the output of $\funF_\theta$ goes through the quantizer $\funQ$ and entropy coder $\funC_\psi$. 

Note that our approach is different from~\cite{Toronto-2021}, where $\vect{t}_k$ contains binary values, and the dimension of $\vect{t}_k$ determines the feedback overhead.
However, we argue that further compression of this feedback is possible and can provide significant gains.
The authors in~\cite{Toronto-2021} also propose an alternative method where $\vect{t}_k$ contains real values that are quantized (using Lloyd's algorithm~\cite{Lloyd}) according to a given overhead budget, and only the BS DNN is further fine-tuned on the quantized inputs.
On the other hand, in our work, the optimization method~\cite{Balle-2018} described above seeks to minimize the feedback entropy (rate) without explicit dependency on the feedback dimensionality. 
Moreover, our approach allows for end-to-end joint training between pilots, users, and BS processing, including the feedback overhead optimization.

\subsection{BS Processing}
\label{sec:BS processing}

During test time, the entropy decoder losslessly reconstructs the received feedback per user, i.e., $\funC_\psi^{-1}(\vect{b}_k) = \vect{\overline{t}}_k$.
During training, the entropy decoder is skipped, and the feedback vectors $\vect{b}_k = \vect{\tilde{t}}_k$ are directly fed into the BS DNN $\funG_\phi$, where $\phi$ represent the set of trainable parameters.
The output of the BS, as defined in~\eqref{eq:BS-output}, formally consists in both the precoding matrix $\mat{V}$ and the reconstructed channel gains $\mat{\hat{H}}$. 
However, in our work we focus on the precoding-oriented output $\mat{V}=\funG_\phi(\vect{\tilde{t}}_1,\dots,\vect{\tilde{t}}_K)$, and  ignore $\mat{\hat{H}}$.
The output of the BS has to satisfy the power constraint by setting $\text{Tr}(\vect{V}\vect{V}^H) = P$.

\subsection{Loss Function}
\label{sec:optimization}

As discussed in Section~\ref{sec:introduction}, three metrics can be considered in the CSI feedback problem: feedback overhead, system performance, and channel distortion.
In order to train the end-to-end system with deep learning techniques, we need a differentiable loss function that emulates the required properties for the system described in Section~\ref{sec:system-model} and depicted in Fig.~\ref{fig:system-model}.
We consider a loss function (to be minimized during training) that combines the three metrics as
\begin{equation}
\label{eq:loss}
    \mathcal{L}(\theta,\phi, \psi) =  \mathcal{O} - \lambda \mathcal{R} + \gamma \mathcal{D},
\end{equation}
where $\mathcal{O}$ represents the feedback overhead, $\mathcal{R}$ represents the system performance, and $\mathcal{D}$ represents the distortion loss; $\lambda$ and $\gamma$ determine the tradeoff between the three components.
We assume that the tradeoff coefficients are non-negative, i.e., $\lambda,\gamma \geq 0$.
For example, traditional overhead-distortion (or rate-distortion) settings correspond to $\lambda=0$, while precoding-oriented systems correspond to $\gamma=0$.
In our work, we will consider $\gamma = 0$ and will sweep values of $\lambda$ for different feedback overhead.
Systems that provide both precoding vectors and channel reconstructions can be also considered by having non-zero values for both $(\lambda,\gamma)$, but they are not the focus of this work.
More details about the metrics are provided below.

\subsubsection{Overhead}
The feedback overhead accounts for the amount of bits that is required to transmit $\vect{b}_k$'s on the uplink.
As discussed previously, we use the empirical entropy of $\vect{\tilde{t}}_k$ as a measure for the feedback overhead~\cite{Balle-2018,ICL-2021}.
Hence, we can express the overhead metric for the $k$-th user as
\begin{equation}
\label{eq:overhead}
    \mathcal{O}_k(\theta, \psi) = \E_{\vect{h}_k,\vect{u}_k, \noisePilots_k} \left[-\log_2 p_{\vect{\tilde{t}}}(\vect{\tilde{t}}_k; \psi)\right],
\end{equation}
where $p_{\vect{\tilde{t}}}(\cdot; \psi)$ represents the approximated density of $\vect{\tilde{t}}$ parameterized by $\psi$.
This loss term can be seen as an estimate for the number of bits required to represent each feedback vector $\vect{b}_k$.
The sum of the feedback overhead for the $K$ users can be expressed as
\begin{equation}
\label{eq:overhead-sum}
    \mathcal{O}(\theta, \psi) = \sum_{k=1}^K \mathcal{O}_k(\theta, \psi).
\end{equation}

\subsubsection{Performance}
The system performance can be evaluated in terms of the achievable rate experienced by the users, as explained in Section~\ref{sec:system-model}.
According to~\eqref{eq:ach-rate-user} and~\eqref{eq:ach-rate-sum}, we recall that the performance metric is 
\begin{equation}
\label{eq:performance-sum}
    \mathcal{R}(\theta, \psi, \phi) = \sum_{k=1}^K \mathcal{R}_k(\theta, \psi, \phi),
\end{equation}
where
\begin{equation}
\label{eq:performance}
    \mathcal{R}_k(\theta, \psi, \phi) = \E_{\vect{h}_k, \mat{U},\NoisePilots} \log_2 \left( 1 + \frac{ |\vect{h}_k^H \vect{v}_k|^2 }{ \sum_{j\neq k } |\vect{h}_k^H \vect{v}_j|^2 + \sigma^2} \right).
\end{equation}
and $\mat{V}=[\vect{v}_1,\dots,\vect{v}_K]=\funG_\phi(\vect{\tilde{t}}_1,\dots,\vect{\tilde{t}}_K)$.

\subsubsection{Distortion}
In order to compare the precoding-oriented approach with conventional methods, we also consider mean squared error (MSE) as a distortion metric.
The distortion component can be expressed as
\begin{equation}
    \mathcal{D}(\theta, \psi, \phi) = \E_{\mat{H}, \mat{U},\NoisePilots} \Norm{\mat{H}-\mat{\hat{H}}}^2_2,
\end{equation}
where $\mat{\hat{H}}=[\vect{\hat{h}}_1,\dots,\vect{\hat{h}}_K]=\funG_\phi(\vect{\tilde{t}}_1,\dots,\vect{\tilde{t}}_K)$ when considering the channel reconstructions at the BS output.

\section{Simulation Results}
\label{sec:results}

While our framework is applicable to any channel model, in order to provide a comparison with the relevant literature~\cite{Toronto-2021}, we consider the following multipath channel model for our simulations.
We assume that the BS is equipped with a uniform linear array, with transmit array response 
\begin{equation}
\label{eq:antenna-resp}
    \vect{a}_t(\beta) = \left[ 1, e^{j \frac{2\pi f_c d}{c} \sin(\beta)}, \dots, e^{j \frac{2\pi f_c d}{c}(N_t -1) \sin(\beta)} \right],
\end{equation}
where $\beta$ denotes the angle of departure (AoD), $d$ denotes the antenna spacing, $f_c$ denotes the carrier frequency and $c$ denotes the speed of light.
The channel gains at the $k$-th user are the summation of $L_p$ propagation paths as 
\begin{equation}
\label{eq:channel-gain}
    \vect{h}_k = \frac{1}{\sqrt{L_p}} \sum_{\ell=1}^{L_p} \alpha_{\ell,k} \vect{a}_t(\beta_{\ell,k}),
\end{equation}
where $\alpha_{\ell,k}$ is the complex gain of the $\ell$-th path between the BS and the $k$-th user.

We focus on the overhead-performance tradeoff in the following analysis.
The system performance is evaluated as the sum achievable rate~\eqref{eq:performance-sum}, while the feedback overhead (per user) is estimated according to~\eqref{eq:overhead}.
For the numerical results presented in this paper, in order to compare with~\cite{Toronto-2021}, we assume that there are $K=2$ users, the channel has $L_p=2$ paths, the pilots' length is $L=8$, the BS is equipped with $N_t=64$ antennas, the transmitted power constraint is $P=1$, and the signal-to-noise ratio (SNR) $P/\sigma^2$ on the received pilots is 10 dB. The feedback $\vect{b}_k$ is assumed to be noiseless.
The main goal of our work is to investigate the precoding-oriented approach, trained for overhead-performance tradeoff, and compare it with conventional channel reconstruction followed by traditional precoding. 

The neural networks hyperparameters are chosen as follows.
The user network $\funF_\theta$ consists of four fully-connected layers with $[1024,2048,256,N_b]$ neurons, where $N_b=16$, while the BS network $\funG_\phi$ has five fully-connected layers with $[1024,512,512,256,K\cdot N_t]$ neurons.
Each hidden layer is preceded by batch normalization, and followed by a rectified linear unit (ReLU) activation.
Real and imaginary parts of signals are processed on separate layers when appropriate.
The last layers of both $\funF_\theta$ and $\funG_\phi$ have linear activation, 
For the overhead mechanism~\cite{Balle-2018}, we use the \emph{entropy bottleneck} class from~\cite{begaint2020compressai}, which provides a PyTorch implementation of~\cite{tfc-github}.
The end-to-end architecture is trained with ADAM optimizer, learning rate $10^{-3}$, over at least $10^6$ batches of size 1024.
The final numerical results are obtained on a test set containing $10^4$ channel realizations for the two users.

\subsection{Baseline algorithms with CSIT}

The best case scenario is when the perfect CSI $\mat{H}$ is available at the BS. 
This case is referred to as CSI at the transmitter (CSIT).
Maximal-ratio transmission (MRT) and zero-forcing (ZF) are two traditional precoding schemes~\cite{MU-beamforming, MIMObook}. 
In MRT, the per-user received power  is maximized, while ZF attempts to minimize the inter-user interference.
Note that in the high (low) SNR asymptotic regime, the optimal linear precoding strategy converges to the ZF (MRT) solution~\cite{MU-beamforming}. 
The precoding matrices for MRT and ZF are 
\begin{align}
\label{eq:MRT}
	\vect{V}_{\text{MRT}} &= \alpha_{\text{MRT}} \vect{H}^H\\
\label{eq:ZF}
	\vect{V}_{\text{ZF}}  &= \alpha_{\text{ZF}} \vect{H}^H (\vect{H} \vect{H}^H)^{-1}
\end{align}
where $\alpha_{\text{MRT}}$ and $\alpha_{\text{ZF}}$ are determined to ensure that the power constraints $\text{Tr}(\vect{V}_{\text{MRT}}\vect{V}_{\text{MRT}}^H) \leq P$ and $\text{Tr}(\vect{V}_{\text{ZF}}\vect{V}_{\text{ZF}}^H) \leq P$ are satisfied.

\subsection{Precoding-oriented System Trained on the Overhead-performance Tradeoff}

As described in this paper, our approach optimizes the overhead-performance tradeoff.
The end-to-end system of Fig.~\ref{fig:system-model} is trained with $\gamma=0$ in~\eqref{eq:loss}, where $\mat{V}$ is the only output at the BS.
We do not consider channel reconstructions $\mat{\hat{H}}$, since the end goal is to optimize performance.
We obtain different working points by changing the value of $\lambda$: large (small) $\lambda$ leads to good (poor) precoding performance with a little (big) feedback overhead.

\subsection{Methods From the Literature}

Conventional methods separate the source coding blocks (compressor and decompressor) from the task block (compute precoding).
Note that previous works~\cite{DL-CSI-FB-2018,DL-CSI-FB-survey-2022,ICL-2021} showed that deep learning-based approaches outperform traditional techniques (e.g., compressed sensing) for the CSI feedback reconstruction problem. 
In particular, \cite{ICL-2021}~showed that a deep learning-based CSI feedback architecture can be successfully trained to optimize the overhead-distortion tradeoff, when using a feedback optimization similar to the one described in Section~\ref{sec:feedback-overhead}; so we will consider the following deep learning-based approach as a surrogate for all conventional methods.
To simulate this reconstruction-oriented approach, according to Fig.~\ref{fig:system-model}, we set $\lambda=0$ in our loss function~\eqref{eq:loss} and consider $\mat{\hat{H}}$ as the output of the BS.  
Then, MRT and ZF precoders are computed using the channel estimates $\mat{\hat{H}}$ according to~\eqref{eq:MRT} and~\eqref{eq:ZF}. 
The resulting precoding matrices are denoted as $\mat{\hat{V}}_\text{MRT}$ and $\mat{\hat{V}}_\text{ZF}$.
The reconstruction $\mat{\hat{H}}$ and the precoding matrices $\mat{\hat{V}}_\text{MRT}$ and $\mat{\hat{V}}_\text{ZF}$ are used to estimate the performance according to~\eqref{eq:performance-sum}.
We train the model for different values of $\gamma$ to obtain neural networks with different overhead-distortion tradeoffs.
For example, large (small) $\gamma$ corresponds to a good (poor) reconstruction with a little (big) feedback overhead. 

We also compare our approach with~\cite{Toronto-2021}.
We consider the best results in~\cite[Fig.~4~and~9]{Toronto-2021}, where the feedback is modeled as a vector of binary values.
Each feedback overhead budget determines the dimension of the feedback in the architecture and the end-to-end system is trained to maximize the sum rate~\eqref{eq:performance-sum}.

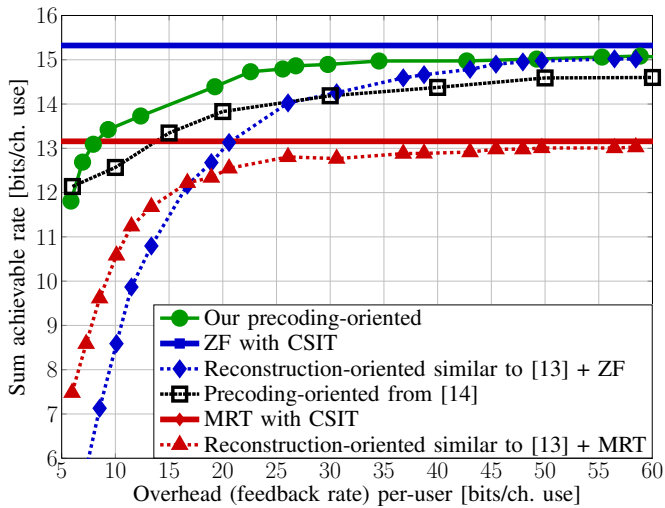
\begin{figure}[t]
    \centering
    \pgfplotsset{
    compat=newest,
    /pgfplots/legend image code/.code={%
        \draw[mark repeat=2,mark phase=2,#1] 
            plot coordinates {
                (0cm,0cm) 
                (0.6cm,0cm)
                (1.2cm,0cm)%
            };
    },
}

\begin{tikzpicture}
\begin{axis}[
	scale only axis,
	font=\huge,
	width=\textwidth,
	height=0.75\textwidth,
	xmajorgrids=true,
	ymajorgrids=true,
	xmin=5,
	xmax=60,
	ymin=6,
	ymax=16,
	xlabel={Overhead (feedback rate) per-user [bits/ch. use]},
	ylabel={Sum achievable rate [bits/ch. use]},
	legend columns=1,
	legend cell align={left},
	legend style={row sep=3pt, at={(1,0)}, anchor = south east},
	]
	
	\addplot+[line width=3pt, \myGreen, solid, mark=*, mark options={scale=3,fill=\myGreen,solid}] table[x=X, y=Y, col sep=comma] 
    {fig/overhead_performance_ours.txt};
	
    \addplot+[line width=5pt, \myBlue, solid] table[x = X, y=Y, col sep=comma] 
    {fig/overhead_performance_ZF_CSIT.txt};
    
    \addplot+[line width=3pt, \myBlue, dashed, mark=diamond*, mark options={scale=3,fill=\myBlue,solid}] table[x = X, y=Y, col sep=comma] 
    {fig/overhead_performance_ZF_CSIR.txt};
    
    \addplot+[line width=3pt, black, densely dotted, mark=square, mark options={scale=3,fill=black,solid}] table[x = X, y=Y, col sep=comma] 
    {fig/overhead_performance_toronto.txt};
    
    \addplot+[line width=5pt, \myRed, solid] table[x = X, y=Y, col sep=comma]
    {fig/overhead_performance_MRT_CSIT.txt};
    
    \addplot+[line width=3pt, \myRed, dashed, mark=triangle*, mark options={scale=3,fill=\myRed,solid}] table[x = X, y=Y, col sep=comma] 
    {fig/overhead_performance_MRT_CSIR.txt};

    \legend{
        Our precoding-oriented,
        ZF with CSIT,
        Reconstruction-oriented similar to~\cite{ICL-2021} + ZF,
        Precoding-oriented from~\cite{Toronto-2021},
        MRT with CSIT,
        Reconstruction-oriented similar to~\cite{ICL-2021}  + MRT
    }
\end{axis}
\end{tikzpicture}
    \caption{
        Analysis of the tradeoff between the feedback overhead and the system performance for $N_t=64$ antennas, $K=2$ users, and $L=8$ pilots.
        Each marker corresponds to a different end-to-end architecture trained for different values of $\lambda$ and $\gamma$ in the loss function~\eqref{eq:loss}.
    }
    \label{fig:main-result}
\end{figure}

\subsection{Comparisons of Results}

We compare our precoding-oriented approach with: MRT/ZF with CSIT; deep learning-based channel reconstruction (reconstruction-oriented system with $\lambda = 0$, as discussed above) followed by MRT/ZF applied to the channel estimates $\mat{\hat{H}}$; the results for the precoding-oriented system from~\cite{Toronto-2021}.

Fig.~\ref{fig:main-result} shows the overhead-performance tradeoff for the above-mentioned methods.
Our precoding-oriented system trained on the overhead-performance tradeoff is shown in green, and it outperforms the other methods for the small feedback overhead regime.
In the large feedback overhead regime, the reconstruction-oriented system (similar to~\cite{ICL-2021}) followed by ZF provides results comparable to our approach, making the two methodologies equivalent when considering the overhead-performance tradeoff. 
Our methods also provide a significant gain in performance compared to~\cite{Toronto-2021} (black line). 
This gain may be explained by the adaptability of our end-to-end solution, which includes a learning mechanism that directly accounts for the feedback overhead in the loss function, as explained in Section~\ref{sec:SemCom-CSI}.
Note that with our method the user's DNN is able to adapt to the overhead budget; i.e., the user is able to learn the efficient precoding-oriented feedback scheme.
As expected, in the large feedback overhead regime, reconstruction followed by MRT/ZF approaches the performance of the CSIT case.
In fact, as the overhead increases, the BS is able to reconstruct the channel with decreasing distortion; hence, the traditional precoding algorithms can rely on more reliable channel estimates.


\section{Conclusion}
\label{sec:conclusion}

In this paper, we have analyzed the tradeoff between feedback overhead, performance, and distortion in the CSI feedback problem for multi-user massive MIMO in FDD.
We showed that the proposed deep learning-based precoding-oriented CSI feedback mechanism provides a higher sum achievable rate than conventional methods for the small overhead regime.
Conventional methods based on channel reconstruction provide equivalent performance to our approach when the system allows for a large overhead.
This work shows the potential of the precoding-oriented CSI feedback mechanism: for a limited overhead budget, the precoding-oriented feedback representation of the channel allows the BS to design better linear precoders to serve the users.
This would potentially allow large massive MIMO arrays to realize their full potential in 6G, in terms of spectral efficiency.

\bibliographystyle{./bibliography/IEEEtran} 
\bibliography{./bibliography/IEEEabrv,
./bibliography/references}

\end{document}